\documentclass[12pt,a4paper]{article}
\usepackage[top=25truemm,bottom=35truemm,left=25truemm,right=25truemm]{geometry}
\usepackage{amsmath,amssymb}
\usepackage{graphicx}

\begin{document}
\setlength{\baselineskip}{18pt}
\begin{titlepage}

\vspace*{1.2cm}
\begin{center}
{\Large\bf Gauge coupling unification in a classically scale invariant model}
\end{center}
\lineskip .75em
\vskip 1.5cm

\begin{center}
{\large Naoyuki Haba}$^1$,
{\large Hiroyuki Ishida}$^1$,
{\large Ryo Takahashi}$^2$, and
{\large Yuya Yamaguchi}$^{1,3}$\\

\vspace{1cm}

$^1${\it Graduate School of Science and Engineering, Shimane University,\\
 Matsue 690-8504, Japan}\\
$^2${\it Graduate School of Science, Tohoku University,\\
 Sendai, 980-8578 Japan}\\
$^3${\it Department of Physics, Faculty of Science, Hokkaido University,\\
 Sapporo 060-0810, Japan}\\

\vspace{10mm}
{\bf Abstract}\\[5mm]
{\parbox{13cm}{\hspace{5mm}
There are a lot of works within a class of classically scale invariant model,
 which is motivated by solving the gauge hierarchy problem.
In this context,
 the Higgs mass vanishes at the UV scale due to the classically scale invariance,
 and is generated via the Coleman-Weinberg mechanism.
Since the mass generation should occur not so far from the electroweak scale,
 we extend the standard model only around the TeV scale.
We construct a model which can achieve the gauge coupling unification at the UV scale.
In the same way, the model can realize
 the vacuum stability, smallness of active neutrino masses, baryon asymmetry of the universe,
 and dark matter relic abundance.
The model predicts the existence vector-like fermions charged under $SU(3)_C$ with masses lower than 1\,TeV,
 and the SM singlet Majorana dark matter with mass lower than 2.6\,TeV.
}}
\end{center}
\end{titlepage}

\section{Introduction}
The Higgs mass parameter $m_h^2$ is only a dimensionful parameter in the standard model (SM),
 and its value is estimated by the observed Higgs mass as
 $\sqrt{-2m_h^2} = M_h =125.09 \pm 0.21 \ ({\rm stat.}) \pm 0.11 \ ({\rm syst.})\,{\rm GeV}$~\cite{Aad:2015zhl}.
Then, a running of the Higgs quartic coupling becomes negative below the Planck scale within the SM.
If the SM can be valid up to a high energy scale such as a breaking scale of a gauge symmetry
 in the grand unification theory (GUT),
 the electroweak (EW) scale should be stabilized against radiative corrections coming from the high energy physics.
To solve the gauge hierarchy problem,
 there are a lot of works motivated by a classically scale invariance~\cite{Hempfling:1996ht}-\cite{Das:2015nwk}. 
The scale invariance prohibits dimensionful parameters at a classical level,
 while it can be radiatively broken by the Coleman-Weinberg (CW) mechanism~\cite{Coleman:1973jx}.
In addition to the classically scale invariance,
 with an additional $U(1)_X$ gauge symmetry, e.g., $U(1)_{B-L}$ gauge symmetry,
 it is possible to naturally realize experimentally observed values of the Higgs mass.
When the $U(1)_X$ symmetry is broken by the CW mechanism,
 the EW symmetry could be also broken through the scalar mixing term.
If the $U(1)_X$ breaking scale is not far from the EW scale,
 the Higgs mass corrections would be sufficiently small,
 and then the hierarchy problem can be solved.
Note that these statements are based on the Bardeen's argument~\cite{Bardeen:1995kv},
 and we consider only logarithmic divergences in this paper
 (see Ref.~\cite{Iso:2012jn} for more detailed discussions).

In this paper, we assume the classically scale invariance at the UV scale,
 where the SM gauge couplings are unified.
We expect that some unknown mechanism, such as a string theory,
 realizes the classically scale invariance and the gauge coupling unification (GCU).
Actually, the GCU can be realized at $3\times 10^{16}\,{\rm GeV}$ in our model,
 and the scale is near the typical string scale ($\sim 10^{17}\,{\rm GeV}$).
To realize the GCU, 
 some additional particles with the SM gauge charges are needed.
Conditions of the GCU can be systematically obtained by an analysis of renormalization group equations (RGEs)~\cite{Giudice:2004tc,Haba:2014oxa}.
When all additional particles are vector-like fermions with the TeV scale masses,
 the GCU scale can be realized between $10^{16}$\,GeV and $10^{17}$\,GeV,
 and there are a lot of possibilities to realize the GCU at the scale.\footnote{
For example, we can consider the origin of the vector-like fermions as the string theory,
 in which a number of vector-like fermions should appear above the compact scale,
 which is expected to be the GCU scale in our model.
Some of them might have the TeV scale masses
 due to the fine-tuning of moduli
 (or Wilson line, extra-dimensional component of anti-symmetric tensor field, and so on).
}
For example, vector-like pairs of quark doublet $Q_{L,R}$ and down-type quark singlet $D_{L,R}$
 can achieve the GCU~\cite{Gogoladze:2010in,Bazzocchi:2012de}.
When there are additional fermions charged under the SM gauge symmetries,
 the gauge couplings and the top Yukawa coupling respectively become larger and smaller compared to the SM case,
 and then, both changes make the $\beta$ function of the Higgs quartic coupling become larger.
Therefore, the vacuum can become stable when the GCU is realized.

To solve the gauge hierarchy problem,
 there should be no intermediate scale between the EW and the GCU scales
 except an energy scale, which is not so far from the EW scale, i.e., the TeV scale.
Then, phenomenological and cosmological problems (e.g.,
 smallness of active neutrino masses, baryon asymmetry of the universe,
 and dark matter (DM))
 should be explained with sufficiently small Higgs mass corrections.
The first two problems can be explained by the right-handed neutrinos,
 which are naturally introduced to cancel the anomalies accompanied with the $U(1)_X$ gauge symmetry,
 via type-I seesaw mechanism~\cite{Minkowski:1977sc}
 and resonant leptogenesis~\cite{Pilaftsis:2003gt}, respectively.
In our model, the DM is identified with the SM singlet Majorana fermions,
 and its stability can be guaranteed by an additional $Z_2$ symmetry~\cite{Okada:2010wd}.
In this paper, we will show that our model can explain the above problems
 as well as realizing the GCU without affecting the hierarchy problem.\footnote{
From theoretical point of view,
 there are some papers constructing a model
 which realizes classically scale invariance and gauge coupling unification at the same scale~\cite{Frampton:1999yb}-\cite{Frampton:2001xh}.
Furthermore,
 asymptotic safety of gravity~\cite{Shaposhnikov:2009pv} leads vanishing couplings at the UV scale,
 which suggests vanishing quartic couplings and gauge coupling unification around the Planck scale
 [see Fig.~1 in Ref.~\cite{Haba:2014qca} for example].
In this paper,
 we simply expect such a situation comes from unknown UV physics.
}

In the next section, we will define our model,
 and explain the $U(1)_X$ gauge symmetry breaking
 as well as the EW symmetry breaking via the CW mechanism. 
We also obtain the upper bound on the $U(1)_X$ breaking scale
 from the naturalness.
In Sec.~\ref{sec:pheno},
 we will discuss the GCU, vacuum stability,
 smallness of active neutrino masses, baryon asymmetry of the universe, and the DM relic abundance.
Our model predicts the existence vector-like fermions charged under $SU(3)_C$ with masses lower than 1\,TeV,
 and the SM singlet Majorana dark matter with mass lower than 2.6\,TeV.
We summarize our results in Sec.~\ref{sec:conclusion}.

\section{Symmetry breaking mechanism}

We consider the $U(1)_X$ gauge extension of the SM
 with three generations of the right-handed neutrinos $\nu_{R_i}$ ($i=1,2,3$),
 six vector-like fermions ($Q_L$, $Q_R$, $D_L$, $D_R$, $N_L$, and $N_R$),
 and two SM singlet scalars ($\Phi$ and $S$).
Charge assignments of the particles are shown in Table\,\ref{tab1}.
The $U(1)_{X}$ charge are given by $B-L+2x_HY$,
 where $x_H$, $B$, $L$, and $Y$ denote
 a real number, the baryon and lepton numbers, and the $U(1)_Y$ hypercharge, respectively.
In particular, $x_H=0$, $-1$ and $-2/5$ correspond to $U(1)_{B-L}$, $U(1)_R$ and $U(1)_\chi$, respectively.
The vector-like fermions $Q_{L,R}$, $D_{L,R}$, and $N_{L,R}$ respectively have the same charges
 as the SM quark doublet, the SM down-quark singlet, and the right-handed neutrino,
 while only the vector-like fermions are odd under an additional $Z_2$ symmetry.
Four of the vector-like fermions ($Q_{L,R}$ and $D_{L,R}$) play a role for achieving the GCU,
 and the others ($N_{L, R}$) are the DM candidates,
 whose stability is guaranteed by the $Z_2$ symmetry. 
These particles are not necessary for the realization of GCU and DM.
We choose them for the simplest extension.

\begin{table}[t]
\begin{center}
\begin{tabular}{|c|c|cc|}\hline
 & $SU(3)_C \otimes SU(2)_L \otimes U(1)_Y$ & $U(1)_{X}$ & $Z_2$ \\
\hline \hline
$q_L$ & (3, 2, 1/6) & $(x_H+1)/3$ & $+$  \\
$u_R$ & (3, 1, $2/3$) & $(4x_H+1)/3$ & $+$  \\
$d_R$ & (3, 1, $-1/3$) & $(-2x_H+1)/3$ & $+$  \\
$\ell_L$ & (1, 2, $-1/2$) & $-x_H-1$ & $+$  \\
$e_R$ & (1, 1, $-1$) & $-2x_H-1$ & $+$  \\
$\nu_R$ & (1, 1, 0) & $-1$ & $+$  \\
$H$ & (1, 2, 1/2) & $x_H$ & $+$  \\
$Q_{L, R}$ & (3, 2, $1/6$) & $(x_H+1)/3$ & $-$  \\
$D_{L, R}$ & (3, 1, $-1/3$) & $(-2x_H+1)/3$ & $-$  \\
$N_{L, R}$ & (1, 1, 0) & $-1$ & $-$  \\
$\Phi$ & (1, 1, 0) & $2$ & $+$ \\
$S$ & (1, 1, 0) & 0 & $+$  \\
\hline
\end{tabular}
\end{center}
\caption{Charge assignment of particles, where $x_H$ is a real number.}
\label{tab1}
\end{table}

The relevant Lagrangian is given by
\begin{eqnarray}
  \mathcal{L} &=& \mathcal{L}_{\rm SM}+\mathcal{L}_{\rm kinetic} - V(H,\Phi,S)
				-( Y_\nu \overline{\ell_L} H^c \nu_R
				+ \kappa_1 \overline{Q_L} H D_R + \kappa_2 \overline{D_L} H Q_R \nonumber \\
			&& 
				+ Y_{M} \Phi \overline{\nu_R^c} \nu_R
				+ Y_{N_L} \Phi \overline{N_L} N_L^c + Y_{N_R} \Phi \overline{N_R^c} N_R \nonumber \\
			&& + f_Q S \overline{Q_L} Q_R + f_D S \overline{D_L} D_R + f_N S \overline{N_L} N_R 
				+ {\rm h.c.}),
\label{Lagrangian}
\end{eqnarray}
 where $\mathcal{L}_{\rm SM}$ is the SM Lagrangian except for the Higgs sector, 
 $\mathcal{L}_{\rm kinetic}$ includes kinetic terms of the Higgs and new particles,
 and $V(H,\Phi,S)$ is a scalar potential of the model.
Without the $Z_2$ symmetry,
 there are also additional Yukawa interactions between the SM particles and the new particles,
 e.g., $y_1 \overline{Q_L} H^c u_R$, $y_2 \overline{Q_L} H d_R$, and $y_3 \overline{q_L} H D_R$.
However, these coupling constants have to be very small
 due to constraints from the precision electroweak data \cite{Barenboim:2001fd}.
To forbid these terms,
 we have imposed odd parity to only the vector-like fermions under the $Z_2$ symmetry.

Since there are two $U(1)$ gauge symmetry,
 $U(1)$ kinetic mixing generally arises in the model.
We can take covariant derivative as
\begin{eqnarray}
	D_\mu = \partial_\mu + i g_3 T^\alpha G_\mu^\alpha + i g_2 T^a W_\mu^a + i g_Y Y B_\mu
			+ i( g_{\rm mix} Y + g_X X ) Z_\mu',
\end{eqnarray}
 where $g$'s are gauge couplings,
 $T^\alpha$ and $T^a$ are generators of $SU(3)_C$ and $SU(2)_L$, respectively,
 and $V_\mu$ ($V=G^\alpha,W^a,B,Z'$) are gauge bosons.
The coupling constant $g_{\rm mix}$ denotes the kinetic mixing
 between the $U(1)_Y$ and the $U(1)_X$ gauge symmetries,
 and we will take $g_{\rm mix}=0$ at the GCU scale.
This boundary condition naturally arises from breaking a simple unified gauge group into
 $SU(3)_C \times SU(2)_L \times U(1)_Y \times U(1)_X$.

We impose the classically scale invariance at the GCU scale,
 and hence, the scalar potential $V(H,\Phi,S)$ is given by
\begin{eqnarray}
	V(H,\Phi,S) = \lambda_H |H|^4 + \lambda_\Phi |\Phi|^4 + \lambda_S S^4
				+ \lambda_{H\Phi} |H|^2 |\Phi|^2 + \lambda_{HS} |H|^2 S^2
				+ \lambda_{\Phi S} |\Phi|^2 S^2,
\label{potential}
\end{eqnarray}
 where there is no dimensionful parameter.
In the model, a complex scalar singlet $\Phi$
 spontaneously breaks the $U(1)_X$ gauge symmetry due to radiative corrections, i.e. the CW mechanism.
Since the complex scalar field obtains the nonzero vacuum expectation value (VEV),
 the SM singlet scalar $\Phi$, the $U(1)_X$ gauge boson $Z'$, the right-handed neutrinos
 and the vector-like fermion $N_{L,R}$ become massive.
After the $U(1)_X$ symmetry breaking,
 negative mass terms of a real scalar singlet $S$ and the SM Higgs doublet $H$ are generated,
 which induces the EW symmetry breaking.
Then, $S$, the vector-like fermions and the SM particles become massive,
 and typically their masses are lighter than those obtained by the $U(1)_X$ symmetry breaking.

Let us explain the symmetry breaking mechanism more explicitly.
We consider the CW potential for a classical field of the singlet scalar $\phi$ as
\begin{eqnarray}
	V_\Phi(\phi) = \frac{1}{4} \lambda_\Phi(v_\Phi)\, \phi^4
						+ \frac{1}{8} \beta_{\lambda_{\Phi}}(v_\Phi)\, \phi^4
							\left( \ln \frac{\phi^2}{v_\Phi^2} -\frac{25}{6} \right),
\label{CWpotential}
\end{eqnarray}
 where we have taken $\Phi = \phi/\sqrt{2}$ without loss of generality,
 and $\langle \phi \rangle= v_\Phi$ is the VEV of $\phi$.
$\beta$ functions of $\Phi$, $\beta_{\lambda_\Phi}$,
 almost depends on quartic terms of $g_X$, $Y_M$ and $Y_{N_{L,R}}$
 for $\lambda_\Phi \simeq 0$.
($\beta$ functions of the model parameters are given in Appendix.)
The effective potential (\ref{CWpotential}) satisfies the following renormalization conditions
\begin{eqnarray}
	\left. \frac{\partial^2 V_\Phi}{\partial \phi^2} \right|_{\phi=0} = 0,\qquad
	\left. \frac{\partial^4 V_\Phi}{\partial \phi^4} \right|_{\phi=v_\Phi} = 6 \lambda_\Phi,
\end{eqnarray}
 and the minimization condition of $V_\Phi$ induces
\begin{eqnarray}
	\lambda_\Phi (v_\Phi) \simeq \frac{11}{6 \pi^2}
		\left[ 6 g_X^4(v_\Phi) - \left( {\rm tr} Y_M^4(v_\Phi) + Y_{N_L}^4(v_\Phi) + Y_{N_R}^4(v_\Phi) \right) \right],
\label{CW_relation}
\end{eqnarray}
 where we have assumed that the scalar quartic couplings are negligibly small in the right-hand side.
When this relation is satisfied,
 the $U(1)_X$ symmetry is broken,
 and $\Phi$ and $Z'$ become massive as
\begin{eqnarray}
	M_\phi = \sqrt{\frac{6}{11} \lambda_\Phi (v_\Phi)} v_\Phi,\qquad
	M_{Z'} = 2 g_X (v_\Phi) v_\Phi,
\label{mass}
\end{eqnarray}
 respectively. 
Since the right-hand side of Eq.\,(\ref{CW_relation}) should be positive,
 $\lambda_\Phi(v_\Phi) \lesssim g_X^4(v_\Phi)$ is required,
 and hence, $M_\phi < M_{Z'}$ is generally expected.
In addition, the quartic terms of Majorana Yukawa couplings ($Y_M$ and $Y_{N_{L,R}}$)
 are smaller than the quartic terms of $g_X$
 because of $\lambda_\Phi(v_\Phi)>0$.
The masses of right-handed neutrinos and $N_{L,R}$ will be discussed in Sec.\,\ref{sec:neutrino}.

After the $U(1)_X$ symmetry breaking,
 the effective potentials for $s$ and $h$ are approximately given by
\begin{eqnarray}
	V_S(s) = \frac{1}{4} \lambda_S s^4 + \frac{1}{4} \lambda_{\Phi S} v_\Phi^2 s^2,\qquad
	V_H(h) = \frac{1}{4} \lambda_H h^4 + \frac{1}{4} \lambda_{H \Phi} v_\Phi^2 h^2,
\label{potential_sh}
\end{eqnarray}
 where $S = s/\sqrt{2}$ and $H=(0, h/\sqrt{2})^T$.
Here, we have assumed that $\lambda_{HS}$ are negligibly small
 compared to $\lambda_{\Phi S}$ and $\lambda_{H\Phi}$ for simplicity.
For $\kappa_{1,2}\simeq 0$, 
 $\lambda_{HS}$ is always negligibly small during renormalization group evolution [see Eq.~(\ref{RGE:HS})].
When $\lambda_{\Phi S}$ and $\lambda_{H \Phi}$ are negative,
 the nonzero VEVs $\langle s \rangle= v_S$ and $\langle h \rangle= v_H$ are obtained as
\begin{eqnarray}
	v_S^2 = \frac{-\lambda_{\Phi S}}{2 \lambda_S} v_\Phi^2,\qquad
	v_H^2 = \frac{-\lambda_{H \Phi}}{2 \lambda_H} v_\Phi^2.
\label{vev}
\end{eqnarray}
Note that $v_S$ and $v_H$ is typically lower than $v_\Phi$,
 because the ratios of quartic couplings
 ($\lambda_{\Phi S}/(2\lambda_S)$ and $\lambda_{H\Phi}/(2\lambda_H)$) should be lower than unity
 to avoid the vacuum instability.
The vector-like fermions and the SM particles become massive,
 while the masses of vector-like fermions ($Q_{L,R}$ and $D_{L,R}$) have to be lower than 1\,TeV
 to realize the GCU as we will show in Sec.\,\ref{sec:GCU}.

In the end of this section,
 we mention the $U(1)_X$ breaking scale, which is described by $v_\Phi$.
Since $M_{Z'}/g_X > 6.9$\,TeV is required from the LEP-II experiments \cite{Carena:2004xs},
 we obtain the lower bound $v_\Phi \gtrsim 3.5$\,TeV.
On the other hand, the naturalness of the Higgs mass suggests a relatively small $v_\Phi$.
A major correction to the Higgs mass is given by $Z'$ intermediating diagrams,
 and one-loop and two-loop corrections are approximately written as
\begin{eqnarray}
	\Delta m_h^2 &\sim& \frac{4 x_H^2 g_X^4 v_\Phi^2}{16\pi^2}\quad {\rm for}\ x_H \neq 0, 
\label{one}\\
	\Delta m_h^2 &\sim& \frac{4 (x_H+1) (4x_H+1)}{9} \frac{y_t^2 g_X^4 v_\Phi^2}{(16 \pi^2)^2},
\label{two}
\end{eqnarray}
 respectively.
When one defines requirement of the naturalness as $\Delta m_h^2 < M_h^2$,
 Eqs.~(\ref{one}) and (\ref{two}) lead the upper bound on $v_\Phi$ as
\begin{eqnarray} 
	v_\Phi &\lesssim& \frac{1}{|x_H|} \left( \frac{0.1}{g_X} \right)^2 \times 10^5 \,{\rm GeV}
		\ {\rm for}\ x_H \neq 0,
\label{vphi1}\\
	v_\Phi &\lesssim& \left( \frac{0.1}{g_X} \right)^2 \times 10^6 \,{\rm GeV},
\label{vphi2}
\end{eqnarray}
 where we have taken $y_t \approx1$.
For $|x_H| < 0.1$,
 the two-loop correction gives stronger bound than one-loop correction.
In the following, we will use the stronger bound for fixed $x_H$.
Note that the mass correction from $\Phi$ is always negligible
 because of a small mixing coupling $\lambda_{H \Phi}$.

\section{Phenomenological and cosmological aspects} \label{sec:pheno}
In this section, we will discuss phenomenological and cosmological aspects of the model:
 the GCU, vacuum stability and triviality, smallness of active neutrino masses,
 baryon asymmetry of the universe, and dark matter.
We will also restrict the model parameters from the naturalness of the Higgs mass.

\subsection{Gauge coupling unification} \label{sec:GCU}
\begin{figure}[t]
\begin{center}
	\includegraphics[scale=1]{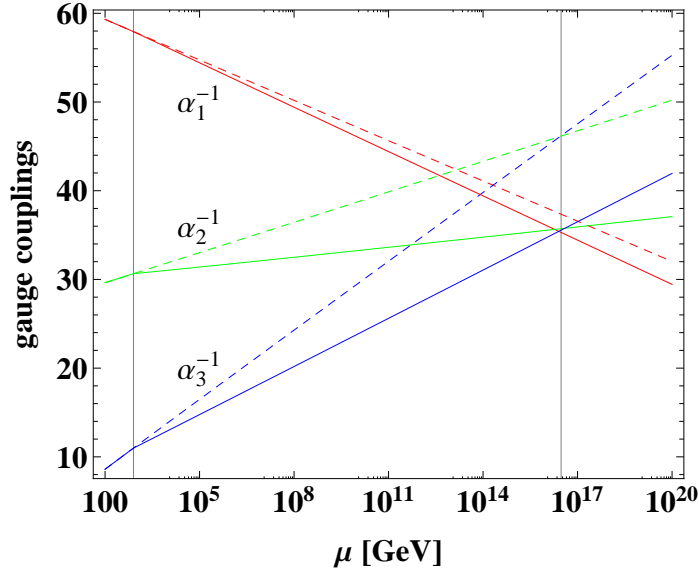}
\end{center}
\caption{Runnings of gauge couplings $\alpha_i^{-1}$.
The dashed and solid lines correspond to the SM and the $U(1)_X$ extended model cases, respectively.
The vertical lines express $M_V=800$\,GeV and $\Lambda_{\rm GCU}=3\times 10^{16}$\,GeV.}
\label{fig:GCU}
\end{figure}

First, we discuss the possibility of the GCU at a high energy scale.
Since four additional vector-like fermions ($Q_{L,R}$ and $D_{L,R}$)
 have gauge charges under the SM gauge groups as shown in Tab.\,\ref{tab1},
 runnings of the SM gauge couplings are modified from the SM.
Then, $\beta$ functions of gauge coupling constants are given by
\begin{eqnarray}
  \beta_{g_Y} = \frac{g_Y^3}{16\pi^2} \frac{15}{2},\qquad
  \beta_{g_2} = \frac{g_2^3}{16\pi^2} \frac{-7}{6},\qquad
  \beta_{g_3} = \frac{g_3^3}{16\pi^2} (-5),
\end{eqnarray}
 at 1-loop level.
Figure \ref{fig:GCU} shows runnings of gauge couplings $\alpha_i^{-1} \equiv 4 \pi /g_i^2$,
 where $U(1)_Y$ gauge coupling is normalized as $g_1\equiv\sqrt{5/3}g_Y$.
The calculation has been done for $x_H=0$ with using 2-loop RGEs.
We note that the running of gauge couplings are almost independent of $x_H$.
In the figure, the horizontal axis is the renormalization scale
 and the vertical axis indicates value of $\alpha_i^{-1}$.
The red, green, and blue lines show $\alpha_1^{-1}$, $\alpha_2^{-1}$, and $\alpha_3^{-1}$, respectively.
The dashed and solid lines correspond to the SM and our model, respectively.
The left vertical line stands for a typical scale of vector-like fermions,
 which has been taken as $M_V=800$\,GeV in Fig.\,\ref{fig:GCU}.
For $\mu<M_V$, the $\beta$ functions are the SM ones,
 and we take boundary conditions for the gauge couplings
 such that experimental values of the Weinberg angle,
 the fine structure constant, and the strong coupling can be reproduced \cite{Buttazzo:2013uya}.
The GCU can be achieved at $\Lambda_{\rm GCU}=(2$--$4)\times10^{16}$\,GeV,
 and the unified gauge coupling is $\alpha_{\rm GCU}^{-1}=(35.4$--$35.8)$.\footnote{
The GCU can be achieved by adjoint fermions as in Ref.~\cite{Haba:2013via,Haba:2013dva}.
}
This is the same result as in Ref.\,\cite{Gogoladze:2010in},
 in which only $Q_{L,R}$ and $D_{L,R}$ are added into the SM.
As the vector-like fermion masses become larger,
 the precision of the GCU becomes worse.
Thus, the masses of $Q_{L,R}$ and $D_{L,R}$ should be lighter than 1\,TeV,
 while vector-like fermion masses are constrained by the LHC experiments
 \cite{CMS:2012ra,Chatrchyan:2013uxa,Aad:2014efa}.
Since the lower bound of vector-like quark lies around 700\,GeV,
 the possibility of the GCU can be testable in the near future.

We note that the proton lifetime in a GUT model.
The proton lifetime is roughly derived from a four-fermion approximation for the decay channel
 $p \rightarrow e^+ + \pi^0$, which is given by
\begin{eqnarray}
	\tau_{\rm p} \sim \left( \alpha_{\rm GCU}^{-1} \right)^2 \frac{\Lambda_{\rm GCU}^4}{m_{\rm p}^5},
\end{eqnarray}
 where $m_{\rm p}$ is the proton mass.
For $\Lambda_{\rm GCU}=3\times10^{16}$\,GeV and $\alpha_{\rm GCU}^{-1}=35.6$,
 we can estimate $\tau_{\rm p} \sim 10^{37}$ yrs,
 which is much longer than the experimental lower bound
 $\tau_{\rm p} > 8.2\times 10^{33}$ yrs \cite{Agashe:2014kda}.
Thus, the model are free from the constraint of the proton decay.

\subsection{Vacuum stability and triviality} \label{sec:vacuum}
Next, we discuss the vacuum stability.
However, it is difficult to investigate exact vacuum stability conditions,
 since there are three scalar fields and each of them has nonzero VEVs.
Therefore, we simply investigate three necessary conditions:
 $\lambda_H>0$, $\lambda_\Phi>0$ and $\lambda_S>0$.

\begin{figure}[t]
\begin{center}
	\includegraphics[scale=0.8]{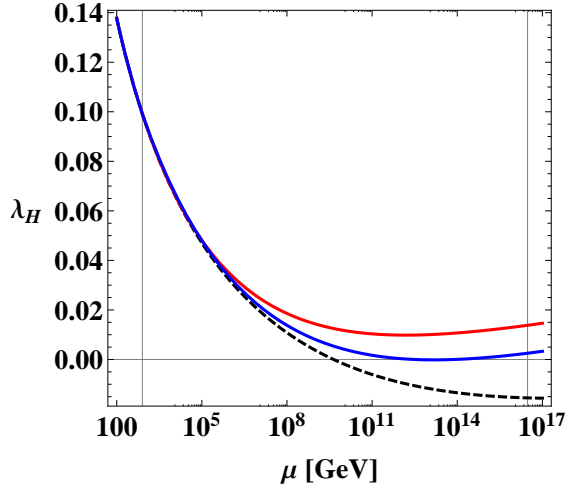}
\end{center}
\caption{Running of $\lambda_H$ in the $U(1)_{B-L}$ ($x_H=0$) case.
The red and blue lines correspond to $\kappa = 0$ and $\kappa = 0.33$, respectively.
The black dashed line shows running of $\lambda_H$ in the SM.
The vertical lines express $M_V=800$\,GeV and $\Lambda_{\rm GCU}=3\times 10^{16}$\,GeV.}
\label{fig:lambda_H}
\end{figure}

The condition $\lambda_H>0$ depends on additional contributions to $\beta_{\lambda_H}$,
 i.e., $\kappa_{1,2}$, $g_X$ and scalar mixing couplings.\footnote{
Running of $\lambda_H$ also depends on mass (or Yukawa coupling) of the top quark.
We will use the central value of world average, i.e., $M_t=173.34\,{\rm GeV}$~\cite{ATLAS:2014wva}.
If we change this value of top quark mass,
 the following numerical results can slightly change.}
If their contributions to $\beta_{\lambda_H}$ are negligible,
 since the SM gauge couplings are larger compared to the SM case,
 running of $\lambda_H$ is raised and always positive.
For example, however,
 the EW vacuum becomes instable for $\kappa \gtrsim 0.33$ in the $U(1)_{B-L}$ ($x_H=0$) case.
We show the running of $\lambda_H$ for $x_H=0$ in Fig.~\ref{fig:lambda_H},
 where $\beta_{\lambda_H}$ is independent of $g_X$ up to the one-loop level,
 and contributions of $g_X$ can be negligible.
The red and blue lines correspond to $\kappa = 0$ and $\kappa = 0.33$, respectively.
The black dashed line shows running of $\lambda_H$ in the SM.
Thus, $\kappa < 0.33$ is required to realize the vacuum stability.

The Higgs mass corrections from $Q_{L,R}$ and $D_{L,R}$ loops are given by
\begin{eqnarray} 
	\Delta m_h^2 \sim \frac{v_S^2}{16\pi^2}
			\left[ (\kappa_1^2 + \kappa_2^2) (f_Q^2 + f_D^2) + 2 \kappa_1 \kappa_2 f_Q f_D \right]
		\sim \frac{12 \kappa^2 M_V^2}{16\pi^2},
\end{eqnarray}
 where we have taken $\kappa = \kappa_1 = \kappa_2$,
 which naturally arises from $L \leftrightarrow R$ symmetry for the vector-like particles,
 and $M_V = M_Q = M_D$ ($M_Q=f_Q v_S/\sqrt{2}$ and $M_D=f_D v_S/\sqrt{2}$) for simplicity.
Then, the naturalness requires $\kappa < 0.1$ for $M_V \sim 1$\,TeV.
Although $\kappa v_H$ is a contribution to the vector-like fermion masses from the Higgs,
 it can be ignored because of $\kappa v_H \ll M_V$.
Since the contribution of $\kappa$ to $\beta_{\lambda_H}$,
 i.e., $24 \lambda_H \kappa^2 - 12 \kappa^4$, is always positive for $\kappa < 0.1$,
 the naturalness condition also guarantees the vacuum stability.
Note that $\kappa\simeq 0$ guarantees $\lambda_{HS}\simeq 0$ at any energy scale,
 which is required to justify our potential analysis for Eq.~(\ref{potential_sh}).

Here, we check contributions of vector-like fermions to the $S$ and $T$ parameters,
 which are approximately given by \cite{Lavoura:1992np,Maekawa:1995ha}
\begin{eqnarray}
	\delta S \approx \frac{43}{30\pi}
		\left( \frac{\kappa v_H}{M_V} \right)^2,\qquad
	\delta T \approx \frac{3(\kappa v_H)^2}{10\pi \sin^2\theta_W M_W^2}
		\left( \frac{\kappa v_H}{M_V} \right)^2,
\end{eqnarray}
 where $\theta_W$ and $M_W$ are the Weinberg angle and the $W$ boson mass, respectively.
For $\kappa < 0.1$, the parameters are estimated as
 $\delta S < 3 \times10^{-4}$ and $\delta T < 2 \times 10^{-5}$,
 which are consistent with the precision EW data $S=0.00 \pm 0.08$ and $T = 0.05 \pm 0.07$
 \cite{Agashe:2014kda}.

The condition $\lambda_\Phi>0$ is almost always satisfied
 when $g_X$ is dominant in the right-hand side of Eq.\,(\ref{CW_relation}),
 i.e., $\lambda_\Phi (v_\Phi) \sim g_X^4(v_\Phi)$.
In this case, $\beta_{\lambda_\Phi}$ is positive up to the GCU scale,
 and then $\lambda_\Phi$ is also positive up to the GCU scale.
It is also possible to realize the critical condition $\lambda_\Phi(\Lambda_{\rm GCU})=0$
 as well as $\lambda_\Phi>0$,
 where the running of $\lambda_\Phi$ is curved upward as in the so-called flatland scenario~\cite{Chun:2013soa,Hashimoto:2013hta,Hashimoto:2014ela,Kawana:2015tka,Haba:2015rha}.
Then, both $g_X$ and Majorana Yukawa couplings are dominant in $\beta_{\lambda_\Phi}$,
 while $\lambda_\Phi$ is much smaller than them.
This means that there is a fine-tuning to satisfy Eq.\,(\ref{CW_relation}).

When $\lambda_S$ is negligible in its $\beta$ function,
 a solution of its RGE is approximately given by
\begin{eqnarray}
	\lambda_S(\mu) \approx \lambda_S(v_S) - 
		\frac{1}{16\pi^2} \left( 12 f_Q^4(v_S) + 6 f_D^4(v_S) + 2 f_N^4(v_S) \right) \ln \frac{\mu}{v_S},
\end{eqnarray}
 where $\mu$ is a renormalization scale.
Once $v_S$ is fixed,
 $f_Q$ and $f_D$ are determined to realize the GCU,
 while $f_N$ remains a free parameter.
To estimate the condition of $\lambda_S>0$,
 we assume $f_N=f_Q=f_D$ at $\mu=v_S$ for simplicity.
Then, we can find that $\lambda_S$ is positive up to the GCU scale
 for $\lambda_S(v_S) \gtrsim 0.01$.
This lower bound of $\lambda_S(v_S)$ is almost unchanged for different values of $v_S$,
 because $v_S$ dependence is logarithmic.

\begin{figure}[t]
\begin{center}
	\includegraphics[scale=0.8]{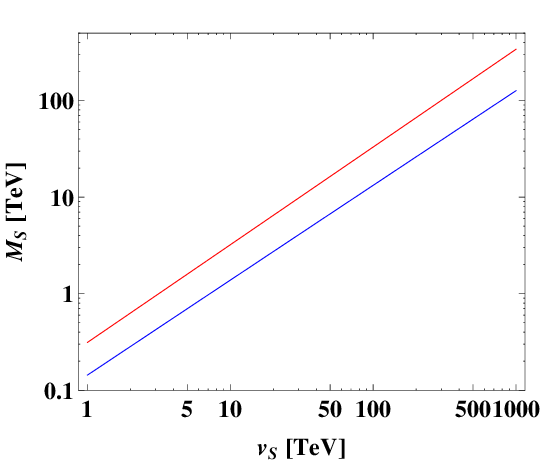}
\end{center}
\caption{$v_S$ dependence on the upper (red) and lower (blue) bounds of $M_s$,
 which correspond to the Landau pole and vacuum stability conditions, respectively.
For the Landau pole bound,
 we take $\Lambda_{\rm LP}=\Lambda_{\rm GCU}=3\times 10^{16}$\,GeV in Eq.\,(\ref{LPs}).}
\label{fig:LPs}
\end{figure}

On the other hand, when $\lambda_S$ is dominant in $\beta_{\lambda_S}$,
 the Landau pole might exist,
 at which the theory is not valid from the point of view of perturbativity (triviality).
The energy scale where the Landau pole appears is approximately estimated as
\begin{eqnarray} 
	\Lambda_{\rm LP} = v_S \exp \left[ \frac{4 \pi^2 v_S^2}{9 M_s^2} \right],
\label{LPs}
\end{eqnarray}
 where $M_s=\sqrt{2\lambda_S(v_S)}v_S$ is a mass of the real singlet scalar field.
Figure \ref{fig:LPs} shows $v_S$ dependence on the upper (red) and lower (blue) bonds of $M_s$,
 which correspond to the Landau pole and vacuum stability conditions, respectively.
Since the both bounds are almost proportional to $v_S$,
 allowed values of $\lambda_S(v_S)$ are almost unchanged for different $v_S$.
We can find a strong constraint for $\lambda_S$ as $0.01 \lesssim \lambda_S(v_S) \lesssim 0.05$.

\begin{figure}[t]
\begin{center}
	\includegraphics[scale=0.6]{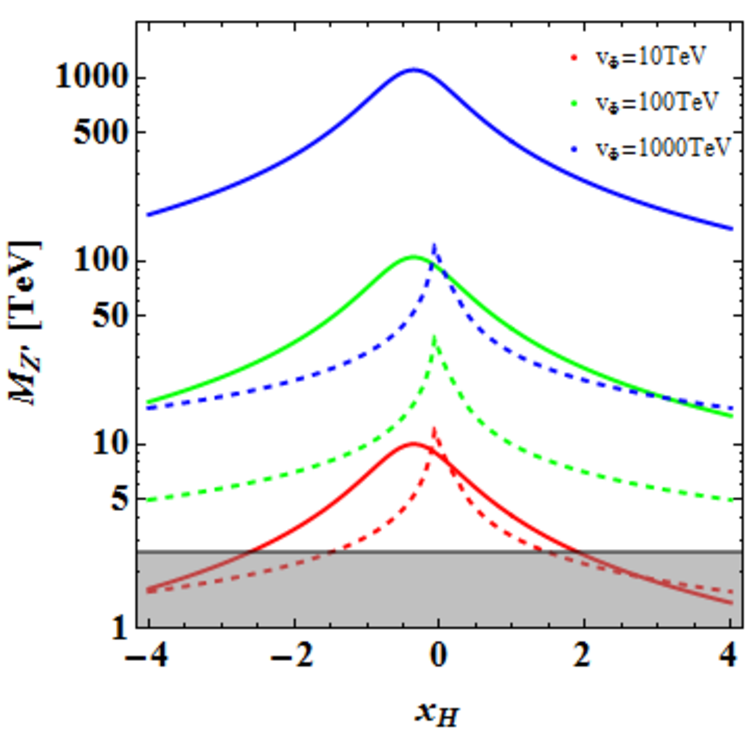}
\end{center}
\caption{The upper bound of $M_{Z'}$ for fixed $v_\Phi$,
 which depends on $x_H$.
The solid and dashed lines show the Landau pole (\ref{LP}) and the naturalness (Eqs.\,(\ref{vphi1}) and (\ref{vphi2})) bounds, respectively.
For the Landau pole bound,
 we take $\Lambda_{\rm LP}=\Lambda_{\rm GCU}=3\times 10^{16}$\,GeV in Eq.\,(\ref{LP}).
The shaded region ($M_{Z'}<2.6$\,TeV) is excluded by the LHC experiments.}
\label{fig:LP}
\end{figure}

In the same way, the Landau pole also exists
 when $g_X(v_\Phi)$ is sufficiently large.
The energy scale where the Landau pole appears is approximately estimated by the one-loop RGE of $g_X$ as
\begin{eqnarray} 
	\Lambda_{\rm LP} = v_\Phi \exp \left[ \frac{32 \pi^2 v_\Phi^2}{(44/3 + 64/3 x_H + 30 x_H^2) M_{Z'}^2} \right],
\label{LP}
\end{eqnarray}
 where $M_{Z'}$ is given in Eq\,(\ref{mass}).
Figure \ref{fig:LP} shows the upper bound of $M_{Z'}$ for fixed $v_\Phi$,
 which depends on $x_H$.
The solid lines show the maximal value of $M_{Z'}$ allowed in the model,
 which are calculated by Eq.\,(\ref{LP})
 for $\Lambda_{\rm LP}=\Lambda_{\rm GCU}=3\times 10^{16}$\,GeV.
Note that the peak of solid lines at $x_H=-16/45$ corresponds to the orthogonal basis of two $U(1)$ gauges.
The dashed lines show the naturalness bound estimated by Eqs.\,(\ref{vphi1}) and (\ref{vphi2}).
The red, green, and blue colors correspond to $v_\Phi=10$, 100, and 1000\,TeV, respectively.
The shaded region ($M_{Z'}<2.6$\,TeV) is excluded by the LHC experiments~\cite{Aad:2014cka,Khachatryan:2014fba}.
When we define the triviality bound as $\Lambda_{\rm GCU} < \Lambda_{\rm LP}$,
 it prohibits the regions above the solid lines.
One can see that the bound leads $g_X(v_\Phi) \lesssim 0.5$ from Eq.\,(\ref{mass}),
 which is almost independent of $v_\Phi$.
Since the naturalness requires the stronger constraints than the triviality bound in almost all parameter space,
 we can say that the naturalness guarantees no Landau pole below the GCU scale.
Note that the both bounds are almost the same for $v_\Phi=10$\,TeV,
 and they exclude $M_{Z'} > 10$\,TeV.

\subsection{Neutrino masses and baryon asymmetry of the universe} \label{sec:neutrino}
From the Lagrangian (\ref{Lagrangian}),
 the neutrino mass terms are given by
\begin{eqnarray}
	(\overline{\nu_L}, \overline{\nu_R^c}, \overline{N_L}, \overline{N_R^c})
	\left( \begin{array}{cccc}
		0 & m_D & 0 & 0 \\
		m_D^T & M_M & 0 & 0 \\
		0 & 0 & M_{N_L} & m_N \\
		0 & 0 & m_N & M_{N_R} 
	\end{array} \right)
	\left( \begin{array}{c}
		\nu_L^c \\ \nu_R \\ N_L^c \\ N_R
	\end{array} \right),
\end{eqnarray}
 where $m_D = Y_\nu v_H / \sqrt{2}$, $M_M = Y_M v_\Phi / \sqrt{2}$,
 $M_{N_{L,R}} = Y_{N_{L,R}} v_\Phi / \sqrt{2}$, and $m_N = f_N v_S / \sqrt{2}$.
There is no mixing term between $\nu_{L,R}$ and $N_{L,R}$ due to the $Z_2$ symmetry.
The active neutrino masses can be obtained by the usual type-I seesaw mechanism \cite{Minkowski:1977sc},
 i.e., $m_\nu \approx m_D M_M^{-1} m_D^T$.
The heavier mass eigenvalue is nearly equal to $M_M$,
 whose upper bound is given by the naturalness of the Higgs mass.
Neutrino one-loop diagram contributes the Higgs mass as
\begin{eqnarray} 
	\Delta m_h^2 \sim \frac{Y_\nu^2 Y_M^2 v_\Phi^2}{16\pi^2}
		\sim \frac{m_\nu M_M^3}{16\pi^2 v_H^2},
\end{eqnarray}
 where we have used the seesaw relation.
For $m_\nu \sim 0.1$\,eV,
 the naturalness requires $M_M \lesssim 10^7$\,GeV.

We mention the baryon asymmetry of the universe.
In the normal thermal leptogenesis~\cite{Fukugita:1986hr},
 there is a lower bound on the right-handed neutrino mass as $M_M \gtrsim 10^9$\,GeV~\cite{Davidson:2002qv}.
However, the resonant leptogenesis can work even at the TeV scale,
 where two right-handed neutrino masses are well-degenerated~\cite{Pilaftsis:2003gt}.
In our model, additional $U(1)_X$ gauge interactions make
 the right-handed neutrinos be in thermal equilibrium with the SM particles~\cite{Iso:2010mv}.
A large efficiency factor can be easily obtained,
 and the sufficient baryon asymmetry of the universe can be generated
 by the right-handed neutrinos with a few TeV masses.
Since the neutrino Yukawa coupling $Y_N$ and $Y_M$ almost do not depend on
 the other phenomenological problems,
 we can do the same analysis as in Ref.~\cite{Iso:2010mv},
 and hence, the result is also the same as in Ref.~\cite{Iso:2010mv}.

For the vector-like neutrinos ($N_{L,R}$),
 we consider $M_N = M_{N_L} = M_{N_R}$,
 which naturally arises from $L \leftrightarrow R$ symmetry for the vector-like fermions.
Then, the mass eigenvalues are respectively $M_{N_1} = |M_N - m_N|$ and $M_{N_2} = |M_N + m_N|$
 for $N_1 = (N_L^c - N_R)/\sqrt{2}$ and $N_2= (N_L^c + N_R)/\sqrt{2}$.
The lighter mass eigenstate $N_1$ is a DM candidate,
 because its stability is guaranteed by the $Z_2$ symmetry.
In the limit of $m_N \to 0$ ($M_{N_1} = M_{N_2}$),
 $N_1$ and $N_2$ are degenerate,
 and $N_2$ is also effective for a calculation of the DM relic abundance.
In the next subsection, we will investigate the degenerate $N_{1,2}$ case.

In our model, the $U(1)_X$ gauge symmetry is successfully achieved via the CW mechanism.
It requires $\lambda_\Phi (v_\Phi)>0$ in Eq.~(\ref{CW_relation}),
 that is,
\begin{eqnarray}
	n_\nu M_M^4 + 2 M_N^4 < \frac{3}{32} M_{Z'}^4,
\label{CW_relation2}
\end{eqnarray}
 where $n_\nu$ is a relevant number of right-handed neutrinos,
 which is defined as ${\rm tr}Y_M^4 (v_\Phi/\sqrt{2})^4 = n_\nu M_M^4$.
Thus, the Majorana masses must be lighter than the $Z'$ boson mass.
We have made sure that this constraint is always satisfied
 when $N_{1,2}$ explain the DM relic abundance.

\subsection{Dark matter} \label{sec:DM}

\begin{figure}[t]
\begin{center}
	\includegraphics[width=15cm]{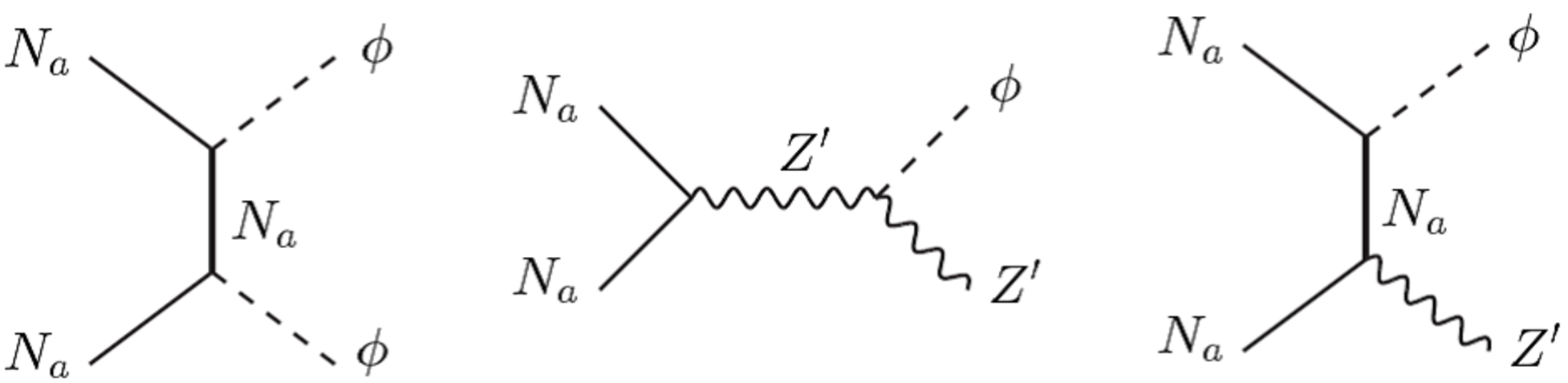}
\end{center}
\caption{Annihilation processes of the dark matter $N_a$ ($a=1,2$).
}
\label{fig:diagram}
\end{figure}

To calculate the DM relic abundance,
 we use the same formula for the DM annihilation cross sections as in Ref.~\cite{Benic:2014aga},
 where a new vector-like fermion is only $N_{L,R}$ (or $N_{1,2}$),
 and the SM fermions do not have $U(1)_X$ charges.
The annihilation processes are
 $t$-channel $NN \to \phi\phi$,
 $t$-channel $NN \to Z'\phi$,
 and  $Z'$ mediated $s$-channel $NN \to Z'\phi$.
The corresponding diagrams are shown in Fig.~\ref{fig:diagram}.
Although our model has other contributions to the annihilation cross sections,
 they are all negligible in the following setup.
We consider the degenerate case for simplicity,
 in which there is no vector-like mass term of $N$.
Thus, $t$-channel $NN \to ss$ process
 and $s$ mediated $s$-channel $NN \to \nu_R \nu_R$ process does not occur at tree level.
From Eq.~(\ref{CW_relation2}), $(2M_N)^2 < M_{Z'}^2$ is always required.
Then, the annihilation cross section $\sigma (NN\to Z^{\prime *} \to f\bar{f})$,
 where $f$ is some $U(1)_X$ charged fermion,
 is suppressed by $1/M_{Z'}^2$.
As a result, we can use the same formula for the DM annihilation cross sections as in Ref.~\cite{Benic:2014aga}.

The spin independent cross section for the direct detection
 is almost dominated by t-channel exchange of scalars $h$ and $\phi$,
 which has been considered in Ref.~\cite{Benic:2014aga}.
However, our model has an additional contribution due to $Z'$ exchange diagrams,
 which is given by~\cite{Duerr:2015wfa}
\begin{eqnarray}
	\sigma_{SI} = \frac{m_n^2 M_N^2}{\pi (m_n+M_N)^2}\frac{g_X^4}{M_{Z'}^4} 
		&=& 7.75 \times 10^{-42}~\left(\frac{\mu_n}{1\,{\rm GeV}}\right)^2
			\left(\frac{1\,{\rm TeV}}{v_\Phi}\right)^4 {\rm cm^2},
\end{eqnarray}
 where $m_n$ is the nucleon mass, and $\mu_n=m_n M_N/(m_n+M_N)$ is the reduced nucleon mass.
For the DM with the masses of 100\,GeV and 1\,TeV,
 the small $v_\Phi$ regions such as $v_\Phi<11\,{\rm TeV}$ and $v_\Phi<6\,{\rm TeV}$
 are excluded by the LUX experiment, respectively~\cite{Akerib:2013tjd}.
These bound are stronger than the LEP bound,
 where $v_\Phi<3.5\,{\rm TeV}$ is excluded.

 In the following, we consider $x_H=0$ ($U(1)_{B-L}$) case.
There are six new parameters in the model:
 the $U(1)_{B-L}$ gauge coupling $g_X$,
 the two Majorana Yukawa coupling $Y_{N_L}$, $Y_{N_R}$,
 the two quartic couplings $\lambda_\Phi$, $\lambda_{H\Phi}$,
 and the VEV of the complex scalar field $v_\Phi$.
On the other hand, there are two conditions $Y_{N_L}=Y_{N_R}$ and Eq.~(\ref{vev}),
 and we require that $N$ explains the DM relic abundance $\Omega_{\rm DM} h^2 = 0.1187$~\cite{Ade:2013zuv}.
Thus, we have three free parameters for the DM analysis.

\begin{figure}[t]
\begin{center}
	\includegraphics[height=7cm]{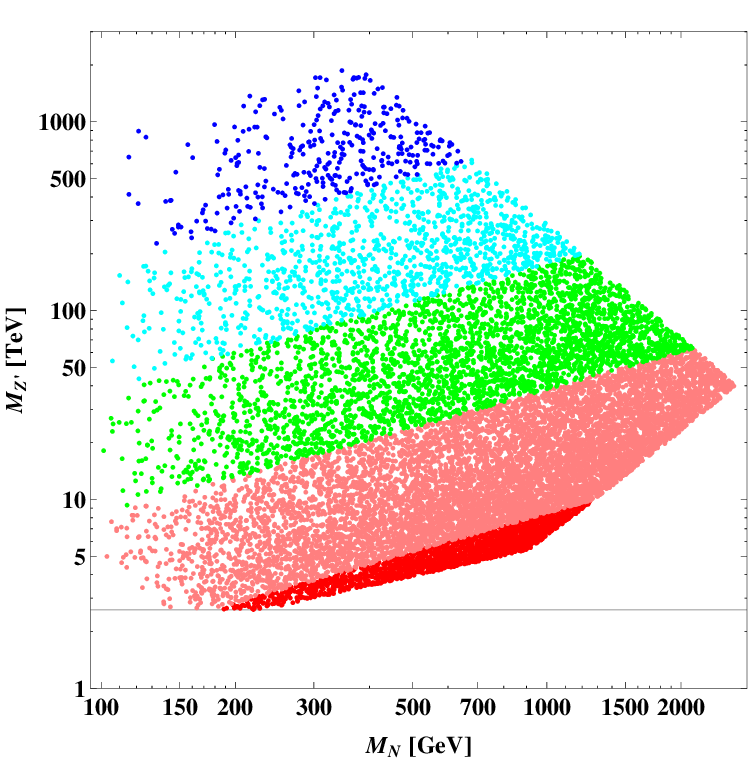} \hspace{5mm}
	\includegraphics[height=7cm]{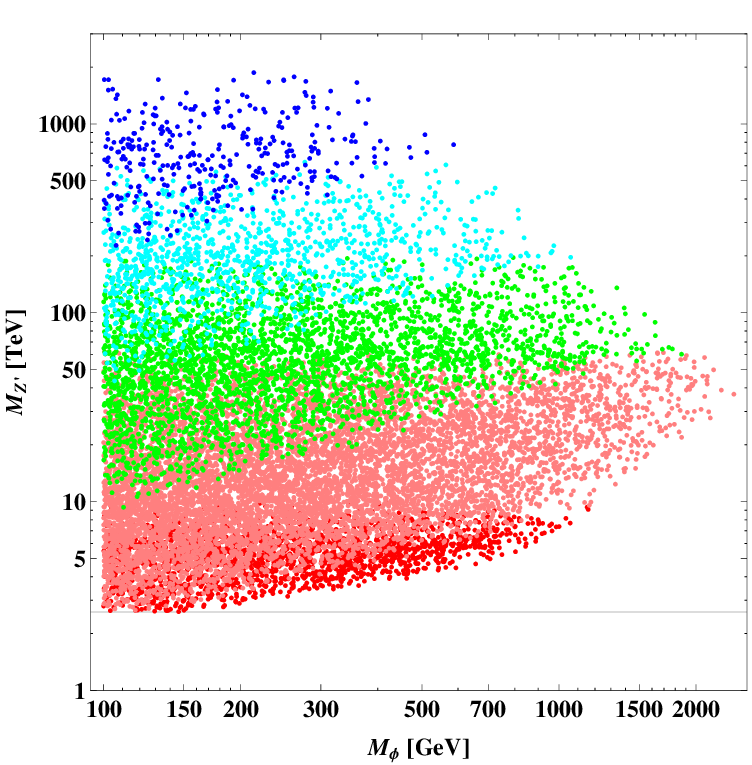}
\end{center}
\caption{Scatter plots in ($M_N$, $M_{Z'}$) plane (left) and ($M_\phi$, $M_{Z'}$) plane (right),
 which realize the DM relic abundance $\Omega_{\rm DM} h^2 = 0.1187$,
 and satisfy all constraints as discussed in this paper as well as the LUX bound.
The horizontal line shows the lower bound on $M_{Z'}$ by the LHC experiments.
The red, pink green, cyan, and blue dots correspond to
 $6\,{\rm TeV} \leq v_\Phi < 10\,{\rm TeV}$,
 $10\,{\rm TeV} \leq v_\Phi < 100\,{\rm TeV}$,
 $100\,{\rm TeV} \leq v_\Phi < 10^3\,{\rm TeV}$,
 $10^3\,{\rm TeV} \leq v_\Phi < 10^4\,{\rm TeV}$,
 $10^4\,{\rm TeV} \leq v_\Phi < 10^5\,{\rm TeV}$, respectively.
}
\label{fig:MN-MZ}
\end{figure}

Figure \ref{fig:MN-MZ} shows scatter plots
 in ($M_N$, $M_{Z'}$) plane (left) and ($M_\phi$, $M_{Z'}$) plane (right),
 which realize the DM relic abundance $\Omega_{\rm DM} h^2 = 0.1187$,
 and satisfy all constraints as discussed above as well as the LUX bound.
The parameter space starts from the initial values
 $M_\phi = 100\,{\rm GeV}$, $M_N = 100\,{\rm GeV}$, and $M_{Z'} = 2.6\,{\rm TeV}$.
Although the two figures in Fig.~\ref{fig:MN-MZ} are very similar,
 $M_N > M_\phi$ is always satisfied.
The region of $M_{Z'} < 2.6\,{\rm TeV}$ is excluded by the current LHC bound~\cite{Aad:2014cka,Khachatryan:2014fba}.
Since $g_X \lesssim 0.5$ is required to avoid the Landau pole,
 the upper bound on $M_{Z'}$ is given by $M_{Z'} \lesssim v_\Phi$,
 while the upper bound in the $M_{N,\phi} \gtrsim 500\,{\rm GeV}$ region is given by the naturalness (\ref{vphi2}).
In the $200\,{\rm GeV} \lesssim M_{N,\phi} \lesssim 900\,{\rm GeV}$ region,
 the lower bound on $M_{Z'}$ is given by the LUX bound.
To realize the DM relic abundance,
 sufficiently large annihilation cross sections are required,
 which induce the lower bound on $M_{Z'}$
 in the $M_N \gtrsim 900\,{\rm GeV}$ region.
From Fig.~\ref{fig:MN-MZ},
 we can see the upper bound on the DM mass as $M_N \lesssim 2.6\,{\rm TeV}$,
 and the bound of $M_\phi$ is almost the same as $M_N$.





\section{Conclusion} \label{sec:conclusion}

To solve the gauge hierarchy problem,
 we have constructed a classically scale invariant model with a $U(1)_X$ gauge extension.
We have assumed the classical scale invariance at the GCU scale,
 where the Higgs mass completely vanishes even with some quantum corrections.
The scale invariance is violated around the TeV scale by the CW mechanism,
 and the Higgs mass can be naturally generated through the scalar mixing term.
The GCU is realized by vector-like fermions $Q_{L,R}$ and $D_{L,R}$,
 which respectively have the same quantum number as the SM quark doublet and down-type quark singlet
 but distinguished by the additional $Z_2$ symmetry,
 and their masses lie in $800\,{\rm GeV} \lesssim M_V \lesssim 1\,{\rm TeV}$.
The GCU scale is $\Lambda_{\rm GCU}=3\times10^{16}$\,GeV with $\alpha_{\rm GCU}^{-1}=35.6$,
 and the proton life time is estimated as $\tau_{\rm p} \sim 10^{37}$ yrs,
 which is much longer than the experimental lower bound $\tau_{\rm p} > 8.2\times 10^{33}$ yrs.

In addition,
 we have shown that the model can explain
 the vacuum stability, smallness of active neutrino masses, baryon asymmetry of the universe,
 and dark matter relic abundance
 without inducing large Higgs mass corrections.
Since there are additional fermions with the SM gauge charges,
 the SM gauge couplings become larger than the SM case,
 which leads smaller top Yukawa couplings.
Then, the $\beta$ function of the Higgs quartic coupling becomes larger,
 and hence the EW vacuum becomes stable.
The smallness of active neutrino masses and the baryon asymmetry of the universe
 can be explained by the right-handed neutrinos
 via the type-I seesaw mechanism and resonant leptogenesis, respectively.
The DM candidate is the SM singlet Majorana fermions $N_{1,2}$,
 and stability of the DM is guaranteed by the additional $Z_2$ symmetry.
We have analyzed the DM relic abundance in the degenerate case ($M_{N_1}=M_{N_2}$),
 and found the upper bound on the DM mass as $M_N \lesssim 2.6\,{\rm TeV}$.

\subsection*{\centering Acknowledgment} \label{Acknowledgement}
The authors thank Y. Kawamura for helpful discussion on the origin of the vector-like fermions.
This work is partially supported by Scientific Grants
  by the Ministry of Education, Culture, Sports, Science and Technology (Nos. 24540272, 26247038, and 15H01037).
The work of Y.Y. is supported
  by Research Fellowships of the Japan Society for the Promotion of Science for Young Scientists
  (Grants No. 26$\cdot$2428).

\section*{Appendix}
\appendix
\section*{{\boldmath $\beta$} functions in the $U(1)_X$ extended SM} \label{app:RGE}
We give one-loop $\beta$-functions in our model:
\begin{eqnarray}
	\beta_{g_Y} &=& \frac{g_Y^3}{16\pi^2} \frac{15}{2},\qquad
	\beta_{g_2} = \frac{g_2^3}{16\pi^2} \frac{-7}{6},\qquad
	\beta_{g_3} = \frac{g_3^3}{16\pi^2} (-5),\\
	\beta_{g_X} &=& \frac{g_X}{16\pi^2} \left[ \left(\frac{44}{3}+\frac{64}{3} x_H+30 x_H^2\right)
		g_X^2+\frac{15}{2} g_{\rm mix}^2+\left(\frac{32}{3}+30 x_H\right) g_{\rm mix} g_X \right], \\
	\beta_{g_{\rm mix}} &=& \frac{1}{16\pi^2} \left[ g_{\rm mix}
		\left( \frac{15}{2} (g_{\rm mix}^2+2 g_Y^2)+\left( \frac{44}{3}+\frac{64}{3} x_H+30 x_H^2 \right)
		g_X^2 \right) \right. \nonumber \\
		&& \left. + \left( \frac{32}{3} +30 x_H\right) g_X (g_Y^2+g_{\rm mix}^2) \right], \\
	\beta_{y_t} &=& \frac{y_t}{16\pi^2} \left[ \frac{9}{2} y_t^2
		- \left( 8 g_3^2 + \frac{9}{4} g_2^2 + \frac{17}{12} (g_Y^2+g_{\rm mix}^2)
		+ \left( \frac{2}{3}+\frac{10}{3} x_H+\frac{17}{3} x_H^2 \right) g_X^2 \right. \right. \nonumber \\
		&& \left. \left. + \left( \frac{5}{3} + \frac{17}{3} x_H \right) g_{\rm mix} g_X \right)
		+ 3 (\kappa_1^2 + \kappa_2^2) \right], \\
	\beta_{Y_M} &=& \frac{Y_M}{16\pi^2} \left[4Y_M^2+2 {\rm tr}Y_M^2+2(Y_{N_L}^2+Y_{N_R}^2)-6 g_X^2\right],\\
	\beta_{Y_{N_L}} &=& \frac{1}{16\pi^2} \left[ Y_{N_L}(6 Y_{N_L}^2+f_N^2+2({\rm tr} Y_M^2+Y_{N_R}^2)
		- 6 g_X^2) + 2 f_N^2 Y_{N_R} \right], \\
	\beta_{Y_{N_R}} &=& \frac{1}{16\pi^2} \left[ Y_{N_R}(6 Y_{N_R}^2+f_N^2+2({\rm tr} Y_M^2+Y_{N_L}^2)
		- 6 g_X^2) + 2 f_N^2 Y_{N_L} \right], \\
	\beta_{\kappa_1} &=& \frac{1}{16\pi^2} \left[ \kappa_1 \left(-8 g_3^2-\frac{9}{4} g_2^2
		- \frac{5}{12} (g_Y^2+g_{\rm mix}^2)+\frac{1}{3} (1-5 x_H)g_{\rm mix} g_X \right. \right. \nonumber\\
		&& \left. \left. + \frac{1}{3}(-2+2 x_H-5 x_H^2)g_X^2 
		+ \frac{1}{2} (f_Q^2+ f_D^2)+3 y_t^2
		+ \frac{9}{2} \kappa_1^2+3 \kappa_2^2 \right)+2 f_Q f_D \kappa_2 \right], \\
	\beta_{\kappa_2} &=& \frac{1}{16\pi^2} \left[ \kappa_2 \left(-8 g_3^2-\frac{9}{4} g_2^2
		- \frac{5}{12} (g_Y^2+g_{\rm mix}^2)+\frac{1}{3} (1-5 x_H)g_{\rm mix} g_X \right. \right. \nonumber\\
		&& \left. \left. + \frac{1}{3}(-2+2 x_H-5 x_H^2)g_X^2 
		+ \frac{1}{2} (f_Q^2+ f_D^2)+3 y_t^2
		+ \frac{9}{2} \kappa_2^2+3 \kappa_1^2 \right)+2 f_Q f_D \kappa_1 \right], \\
	\beta_{f_Q} &=& \frac{1}{16\pi^2} \left[ f_Q \left(-8 g_3^2-\frac{9}{2} g_2^2
		- \frac{1}{6} (g_Y^2+g_{\rm mix}^2)-\frac{2}{3} (1+x_H)g_{\rm mix} g_X \right. \right. \nonumber \\
		&& \left. \left. - \frac{2}{3} (1+x_H)^2 g_X^2 
		+ \frac{1}{2} (\kappa_1^2+\kappa_2^2)+15 f_Q^2+6 f_D^2+2 f_N^2 \right)
		+ 2 f_D \kappa_1 \kappa_2 \right], \\
	\beta_{f_D} &=& \frac{1}{16\pi^2} \left[ f_D \left( -8 g_3^2-\frac{2}{3} (g_Y^2+g_{\rm mix}^2)
		+ \frac{4}{3} (1-2 x_H)g_{\rm mix} g_X-\frac{2}{3} (1-2 x_H)^2 g_X^2 \right. \right. \nonumber \\
		&& \left. \left. + (\kappa_1^2+\kappa_2^2)
		+ 12 f_Q^2+9 f_D^2+2 f_N^2 \right) + 4 f_Q \kappa_1 \kappa_2 \right], \\
	\beta_{f_N} &=& \frac{f_N}{16\pi^2} \left[ -6 g_X^2+2 Y_{N_L}^2+8 Y_{N_L} Y_{N_R}+2 Y_{N_R}^2
		+ 12 f_Q^2+6 f_D^2+5 f_N^2 \right],
\end{eqnarray}

\begin{eqnarray}
	\beta_{\lambda_H} &=& \frac{1}{16\pi^2} \left[ 24\lambda_H^2+\lambda_{H\Phi}^2+2 \lambda_{HS}^2
		+ \lambda_H ( 12y_t^2-9 g_2^2-3(g_Y^2+g_{\rm mix}^2)-12 x_H^2 g_X^2 \right. \nonumber \\
		&& \left. - 12 x_H g_{\rm mix} g_X+12(\kappa_1^2+\kappa_2^2) ) -6 y_t^4-6(\kappa_1^4+\kappa_2^4)
		+ \frac{3}{8} \left\{ 2g_2^4+(g_2^2+(g_Y^2+g_{\rm mix}^2))^2\right\} \right. \nonumber \\
		&& \left. + 3 x_H g_{\rm mix} g_X(g_2^2+g_Y^2+g_{\rm mix}^2+4 x_H^2 g_X^2)
		+ 3 x_H^2 g_X^2 (g_2^2+g_Y^2+3 g_{\rm mix}^2+2 x_H^2 g_X^2) \right], \\
	\beta_{\lambda_\Phi} &=& \frac{1}{16\pi^2} \left[ 20 \lambda_\Phi^2+2 \lambda_{H\Phi}^2
		+ 2 \lambda_{\Phi S}^2+\lambda_\Phi(8({\rm tr} Y_M^2+Y_{N_L}^2+Y_{N_R}^2)-48 g_X^2)
		+ 96 g_X^4 \right. \nonumber \\ 
		&& \left. - 16 ({\rm tr} Y_M^4+Y_{N_L}^4+Y_{N_R}^4) \right], \\
	\beta_{\lambda_S} &=& \frac{1}{16\pi^2} \left[ 72 \lambda_S^2+2 \lambda_{HS}^2+\lambda_{\Phi S}^2
		+\lambda_S(48 f_Q^2+24 f_D^2+8 f_N^2)-12 f_Q^4-6 f_D^4-2 f_N^4 \right], \\
	\beta_{\lambda_{H\Phi}} &=& \frac{1}{16\pi^2} \left[4 \lambda_{HS} \lambda_{\Phi S}
		+ \lambda_{H\Phi} \left( 12 \lambda_H+8 \lambda_\Phi+4 \lambda_{H\Phi}-\frac{9}{2} g_2^2
		- \frac{3}{2} (g_Y^2+g_{\rm mix}^2) \right. \right. \nonumber \\
		&& \left. \left. - 6(4+x_H^2)g_X^2-6 x_H g_{\rm mix} g_X+6 y_t^2+6(\kappa_1^2+\kappa_2^2)
		+ 4({\rm tr} Y_M^2+Y_{N_L}^2+Y_{N_R}^2) \frac{}{} \right) \right. \nonumber \\
		&& \left. + 12 g_X^2 (g_{\rm mix} + 2 x_H g_X)^2 \right], \\
	\beta_{\lambda_{HS}} &=& \frac{1}{16\pi^2} \left[ 2 \lambda_{H\Phi} \lambda_{\Phi S}
		+ \lambda_{HS} \left( 12 \lambda_H+24 \lambda_S+8 \lambda_{HS}-\frac{9}{2} g_2^2
		- \frac{3}{2} (g_Y^2+g_{\rm mix}^2) \right. \right. \nonumber 
\label{RGE:HS}\\
		&& \left. \left. -6 x_H^2 g_X^2-6 x_H g_{\rm mix} g_X+6 y_t^2+6(\kappa_1^2+\kappa_2^2)
		+ 24 f_Q^2+12 f_D^2+4 f_N^2 \right) \right. \nonumber \\
		&& \left. - 12(f_Q^2+f_D^2)(\kappa_1^2+\kappa_2^2)-24 f_Q f_D \kappa_1 \kappa_2 \right], \\
	\beta_{\lambda_{\Phi S}} &=& \frac{1}{16\pi^2} \left[ 4 \lambda_{HS} \lambda_{H\Phi}
		+ \lambda_{\Phi S} \left( 24 \lambda_S+8 \lambda_\Phi+8 \lambda_{\Phi S}-24 g_X^2
		+ 24 f_Q^2+12 f_D^2+4 f_N^2 \right. \right. \nonumber \\
		&& \left. \left. + 4({\rm tr} Y_M^2+Y_{N_L}^2+Y_{N_R}^2) \right)
		- 16 f_N^2 (Y_{N_L}^2+Y_{N_R}^2+Y_{N_L} Y_{N_R}) \right]. 
\end{eqnarray}



\end{document}